\documentclass[twocolumn,aps,superscriptaddress,showpacs,floatfix]{revtex4-1}
\usepackage{amssymb}
\usepackage{amsmath}
\usepackage{graphicx}
\usepackage[normalem]{ulem}
\usepackage[dvips]{color}
\usepackage{bm}
\usepackage{longtable}
\usepackage{array}
\usepackage{etoolbox}
\usepackage{setspace}
\usepackage[colorlinks,linkcolor=blue,anchorcolor=blue,citecolor=blue,urlcolor=blue]{hyperref}

\renewcommand\sout{\bgroup \color{red} \ULdepth=-.5ex \ULset}

\begin{document}

\title{Systematic study of favored $\alpha$-decay half-lives of closed shell odd-$A$ and doubly-odd nuclei related to ground and isomeric states}

\author{Xiao-Dong Sun}
\affiliation{School of Math and Physics, University of South China, 421001 Hengyang, People's Republic of China}
\author{Ping Guo}
\affiliation{School of Math and Physics, University of South China, 421001 Hengyang, People's Republic of China}
\author{Xiao-Hua Li}
\email{lixiaohuaphysics@126.com }
\affiliation{School of Nuclear Science and Technology, University of South China, 421001 Hengyang, People's Republic of China}
\affiliation{Cooperative Innovation Center for Nuclear Fuel Cycle Technology $\&$ Equipment, University of South China, 421001 Hengyang, People's Republic of China}


\begin{abstract}

In this work, we systematically investigate the favored $\alpha$-decay half-lives and $\alpha$ preformation probabilities of both odd-$A$ and doubly-odd nuclei related to ground and isomeric states around the doubly magic cores at $Z=82$, $N=82$ and at $Z=82$, $N=126$, respectively, within a two-potential approach from the view of the valence nucleon (or hole). The results show that the $\alpha$ preformation probability is linear related to $N_\text{p}N_\text{n}$ or $N_\text{p}N_\text{n}I$, where $N_\text{p}$, $N_\text{n}$, and $I$ are the number of valence protons (or holes), the number of valence neutrons (or holes), and the isospin of the parent nucleus, respectively. Fitting the $\alpha$ preformation probabilities data extracted from the differences between experimental data and calculated half-lives without a shell correction, we give two analytic formulas of the $\alpha$ preformation probabilities and the values of corresponding parameters. Using those formulas and the parameters, we calculate the $\alpha$-decay half-lives for those nuclei. The calculated results can well reproduce the experimental data.
\end{abstract}

\maketitle

\section{Introduction}

During the past three decades, many exotic nuclei and decay modes have been found with the advent of radioactive ion beam facilities at GSI, Berkeley, Dubna, Grand Accelerateur National d'Ions Lourds (GANIL), Rikagaku Kenkyusho (RIKEN), and (Heavy Ion Research Facility in Lanzhou) HIRFL~\cite{Gee06,Hof00,Pfu12,And13,Kal13,Ma15,Yan15}. $\alpha$ decay, as one of main decay modes of heavy and superheavy nuclei, attracts constant attention~\cite{Khu14,Car14,Del15,Ni15}. Theoretically, $\alpha$ decay shares the similar theory of barrier penetration with different kinds of charged particles radioactivity, such as single proton emission, heavy ion emission, spontaneous fission~\cite{Buc91,Del09,Poe11,Don09,San12,Bao15} and so on. Experimentally, $\alpha$ decay spectroscopy of very neutron-deficient nuclei and heavy and superheavy nuclei provides much unique structure information. Some spectra can not be described within the mean-field assumption, which indicates the existence of an $\alpha$ cluster in $^{212}$Po~\cite{Ast10}. The three lowest states in energy spectrum of $^{186}$Pb, corresponding to spherical, oblate and prolate shape, are produced by $\alpha$ decay of $^{190}$Po~\cite{And00}. 
Meanwhile, the $\alpha$-decay process is also important for understanding such crucial problems in stellar nucleosynthesis~\cite{Fyn05}, the chronology of the solar system~\cite{Kin12}, and nuclear clustering structure in heavy and superheavy nuclei~\cite{Toh01,Del04,Kar06}.

For nuclei above the shell closures, an enhancement of $\alpha$ decay energies has been recognized through atomic mass data. Thus, an island of the $\alpha$ radioactive nuclei arises above the magic numbers $Z=50$, $N=50$~\cite{Mac65,Pat16}. However, smaller $\alpha$ preformation probabilities hinder $\alpha$ decay of closed shell nuclei, due to the limited number of valence nucleons. Up to now, the formation of $\alpha$-like four-nucleon correlations in nuclear systems has been understood to some extent. For an $\alpha$ decaying state, there may be two special neutrons and protons, which are different from other nucleons moving in the mean field, eventually constituting the $\alpha$ cluster~\cite{Kar06}. Microscopically, $\alpha$ preformation probabilities can be evaluated through an overlap between the initial state and the $\alpha$ decaying state~\cite{Lov98}. For the difficulties coming from a complex quantum many-body system and nuclear force which has still not been pinned down to exact form, the results of $\alpha$-decay widths by using a microscopic calculation applying the Skyrme interaction and the $R$-matrix formulation suggest the missing effects, such as pairing correlations and other residual interactions~\cite{War13}. The approach of the Tohsaki-Horiuchi-Schuck-R\"opke wave function was used to calculate the $\alpha$ preformation probability~\cite{Rop14,Xu16}, which was also successful in describing the cluster structure in light nuclei. The cluster-formation model was also another convenient way to estimate the $\alpha$ preformation factors through the formation energy of an $\alpha$ cluster~\cite{Den16}. But systematic and microscopic calculations of an $\alpha$ cluster preformation are still inaccessible. Therefore, $\alpha$ preformation probabilities are usually extracted from rates of experimental $\alpha$ decay half-lives to theoretical results calculated without considering the preformation factors~\cite{Zha11,Qia13,Zha09}. Recently a semi-empirical formula of $\alpha$ preformation probabilities taking into account the shell effect, the pairing effect, and the angular momentum effect has been given~\cite{Sei15}. Seif \textit{et al.} have proposed that the $\alpha$ preformation probability is proportional to $N_\text{p}N_\text{n}$ for even-even nuclei around proton $Z=82$, neutron $N=82$ and 126 shell closures~\cite{Sei11}. However, it is interesting to test whether odd-$A$ and doubly-odd nuclei also satisfy this relationship or not. In this work, we systematically study favored $\alpha$ decay half-lives and $\alpha$ preformation probabilities of both odd-$A$ and doubly-odd nuclei related to ground and isomeric states around the doubly magic cores at $Z=82$, $N=82$ and at $Z=82$, $N=126$ shell closures, respectively. A good linear relationship between the $\alpha$ preformation probability and $N_\text{p}N_\text{n}$ is found, which shows the importance of valence proton-neutron correlation on the $\alpha$ preformation probabilities. The calculated $\alpha$-decay half-lives can well reproduce the experimental data.

This article is organized as follows. In Section II, the theoretical framework for the calculation of the $\alpha$-decay half-lives is briefly described. The results and discussions are given in Section III. In this section, at first we compare the $\alpha$ preformation probabilities of nuclear isomers to those of ground states, and then the $\alpha$ preformation probabilities are analyzed from the view of the valence proton-neutron interaction.  Sec. IV is a brief summary.

\section{Theoretical framework}

In the framework of Gamow's theory, $\alpha$ decay is described as a preformed $\alpha$ particle moving in the decaying nucleus until the $\alpha$ particle penetrates Coulomb barrier~\cite{Gam28}. $\alpha$-decay half-lives can be calculated as
\begin{eqnarray}\label{1}
T_{1/2}=\frac{\hbar \ln2}{\Gamma},
\end{eqnarray}
where $\Gamma$ is the $\alpha$ decay width. The range of $\Gamma$ is about $10^{-14}$ to $10^{-46}$ MeV, much smaller than the $\alpha$-decay energy $Q_{\alpha}$ near $10$ MeV. Therefore, it is appropriate to treat $\alpha$ decay as a stationary state problem. The two-potential approach has been proposed to deal with metastable states, widely used to calculate $\alpha$ decay half-lives $T_{1/2}$~\cite{Gur87,Qia12,*Qia11,*Qian11}. In this framework, the $\alpha$ decay width can be written as
\begin{eqnarray}\label{2}
\Gamma=P_\alpha \frac{F}{4\mu} \exp[-2\int_{r_2}^{r_3}k(r)\mathit{d}r],
\end{eqnarray}
where $\mu$ is the reduced mass between the $\alpha$ particle and daughter nucleus. $k(r)=\sqrt{\frac{2\mu}{\hbar^2}\mid Q_\alpha-V(r)\mid}$ is the wave number of the $\alpha$ particle, and $r$ is the mass center distance between the preformed $\alpha$ particle and the daughter nucleus. $V(r)$ is the total $\alpha$-core potential. The last exponential term in Eq.~(\ref{2}) is the semiclassical Wentzel-Kramers-Brillouin (WKB) barrier penetrability probability, often called the Gamow factor. $F$ is the normalized factor, describing the $\alpha$ particle assault frequency, which can be approximately given by
\begin{eqnarray}\label{3}
F\int_{r_1}^{r_2}\frac{\mathit{d}r}{2k(r)}=1,
\end{eqnarray}
where $r_1$, $r_2$, and $r_3$ are the classical turning points which satisfy conditions $V(r_1)=V(r_2)=V(r_3)=Q_\alpha$.

$P_\alpha$ is the $\alpha$ preformation probability, which abruptly decreases near the nuclear shell closures and varies smoothly in the region of an open shell~\cite{Qia13,Zha11,Zha09}. With the increasing number of valence nucleons away from the shell closure, the $\alpha$ preformation probability increases. Until close to the next shell closure, the $\alpha$ preformation probability decreases with the decreasing number of valence holes. Regarding this picture, we have completed the systematic calculations of $\alpha$ decay half-lives with shell correction based on the parabola approximation of $\alpha$ preformation probabilities between the neighboring shell closures~\cite{Sun16}. Actually, $P_\alpha$ can be extracted from ratios of calculated $\alpha$ decay half-lives $T^\text{calc}_{1/2}$ to experimental data $T^\text{expt}_{1/2}$, which is defined as $P_\alpha=P_0T^\text{calc}_{1/2}/T^\text{expt}_{1/2}$. The calculated half-lives are obtained with the assumption that $\alpha$ preformation probabilities keep constant for one certain kind of nuclei, such as even-even nuclei, odd-A nuclei, or doubly-odd nuclei. According to the calculations by using the density-dependent cluster model~\cite{Xu05}, the constant factor of preformation probability $P_0$ is taken as $P_0=0.43$ for even-even nuclei, $P_0=0.35$ for odd-A nuclei, and $P_0=0.18$ for doubly-odd nuclei. For odd-A and doubly-odd nuclei, unpaired nucleons result in smaller $\alpha$ preformation probabilities than even-even nuclei due to the block effect.

The total $\alpha$-core potential $V(r)$, including nuclear, Coulomb, and centrifugal
potentials, which is critical for the calculations of $\alpha$-decay widths, can be expressed as
\begin{eqnarray}\label{4}
V(r)=V_\text{N}(r)+V_\text{C}(r)+V_\text{l}(r),
\end{eqnarray}
where $V_\text{N}(r)$, $V_\text{C}(r)$, and $V_\text{l}(r)$ represent the nuclear, Coulomb, and centrifugal potentials, respectively. In this work, we choose a type of cosh parametrized form for the nuclear
potential, obtained by analyzing experimental data of $\alpha$ decay~\cite{Buc90}, which can be expressed as
\begin{eqnarray}\label{5}
V_\text{N}(r)=-V_0\frac{1+\cosh(R/a)}{\cosh(r/a)+\cosh(R/a)},
\end{eqnarray}
where $V_0$ and $a$ are the depth and diffuseness for the nuclear potential, respectively. In our previous work, we have analysed the experimental $\alpha$-decay half-lives of 164 even-even nuclei to obtain a set of parameters, which is $a=0.5958$~fm and $V_0=192.42+31.059\frac{N-Z}{A}$~MeV~\cite{Sun16}, where $N$, $Z$, and $A$ are the neutron, proton, and mass numbers of the daughter nucleus, respectively. In this work, the parameters of diffuseness and depth for the nuclear potential remain unchanged. $V_\text{C}(r)$ is the Coulomb potential and is taken as the potential of a uniformly charged sphere with sharp radius $R$, which can be expressed as
\begin{eqnarray}\label{6}
V_\text{C}(r)= \left \{
\begin{aligned}
\frac{Z_\text{d}Z_{\alpha}\mathit{e}^2}{2R}[3-(\frac{r}{R})^2],~~~r<R,\\
\frac{Z_\text{d}Z_{\alpha}\mathit{e}^2}{r},~~~~~~~~~~~~~~~~r>R,
\end{aligned}
\right.
\end{eqnarray}
where $Z_\text{d}$ and $Z_\alpha$ are the number of protons in the daughter nucleus and the preformed $\alpha$ particle, respectively. The last part in the $V(r)$, the centrifugal potential, can be estimated by
\begin{eqnarray}\label{7}
V_\text{l}(r)=\frac{l(l+1) \hbar^2}{2\mu r^2},
\end{eqnarray}
where $l$ is the orbital angular momentum taken away by the $\alpha$ particle. $l=0$ for the favored $\alpha$ decays, while $l\neq 0$ for the unfavored decays. The sharp radius $R$ is calculated by the following relationship
\begin{eqnarray}\label{8}
R=1.28A^{1/3}-0.76+0.8A^{-1/3}.
\end{eqnarray}
This empirical radius formula, which is derived from the nuclear droplet model and proximity energy, is commonly used to calculate $\alpha$-decay half-lives~\cite{Roy00}.

\section{Results and discussions}
The aim of this work is to study the $\alpha$-decay half-lives and $\alpha$ preformation probabilities of both odd-A and doubly-odd nuclei around the shell closures taking into account the valence proton-neutron interaction. For the odd-$A$ and doubly-odd nuclei, there may be some excitations of a single nucleon, which can bring about nuclear high-spin isomers in terms of the shell model, and these isomers are similar to ground states with regard to $\alpha$ decay. Thus both ground and  isomeric states should be considered as candidates for $\alpha$-decay parent and daughter nuclei in a unified way~\cite{Ni10,Qia12,*Qia11,*Qian11}. In this case, $\alpha$ transitions can be divided into four kinds, i.e. from ground state to ground state (g.s. to g.s.), from ground state to isomeric state (g.s. to i.s.), from isomeric state to ground state (i.s. to g.s.), and from isomeric state to isomeric state (i.s. to i.s.). Especially, $\alpha$ decays that belong to odd-$A$ and doubly-odd nuclei around a closed shell become even more complicated than even-even nuclei due to the unpaired nucleon.

\begin{figure}[!t]
\centering
\includegraphics[width=8.6cm]{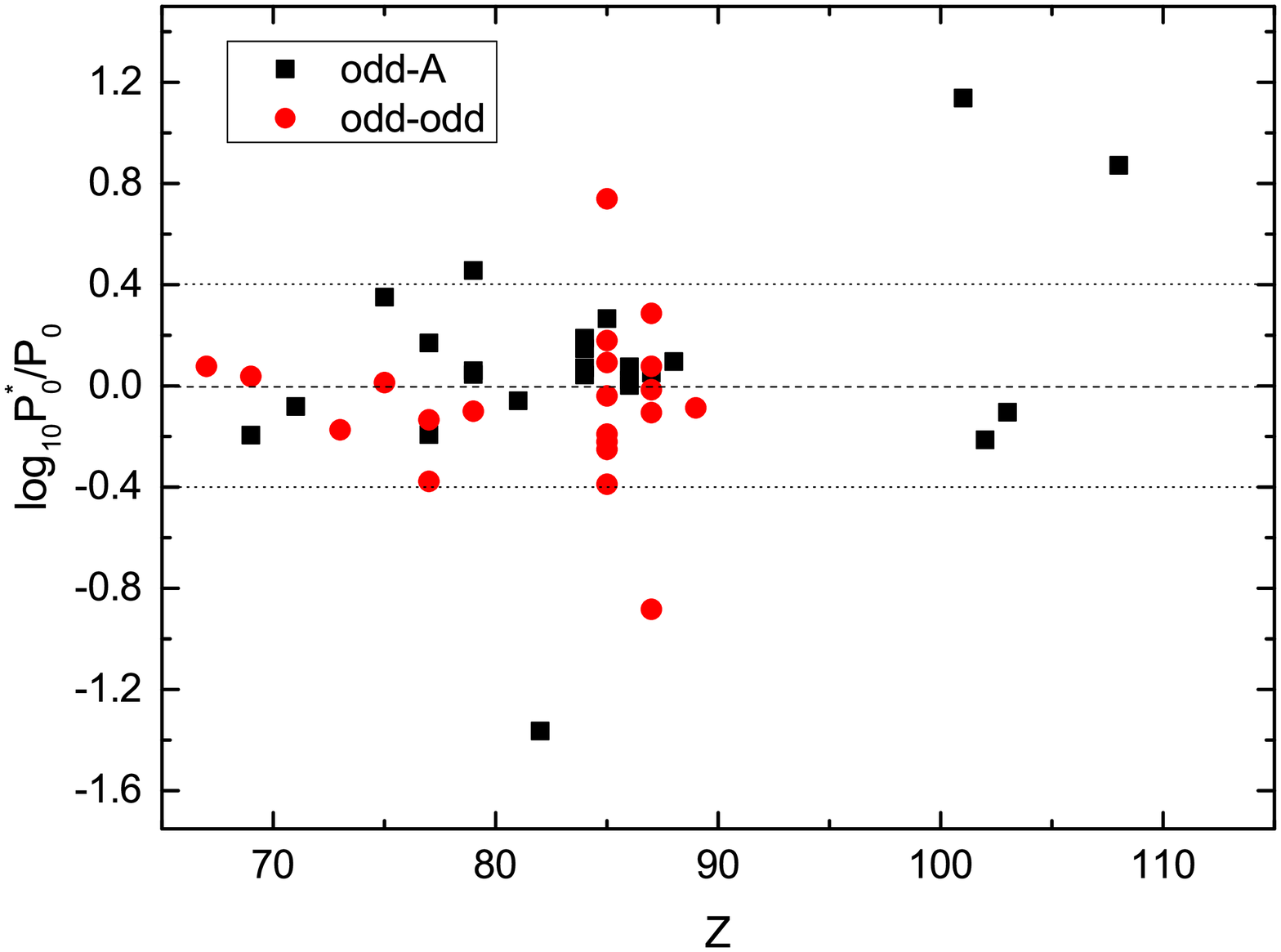}
\caption{\label{Fig1} Logarithm of the $\alpha$ preformation probabilities ratio $P^*_\alpha/P_\alpha$ as a function of proton numbers of the parent nuclei. The $P^*_\alpha$ and $P_\alpha$ denote the $\alpha$ preformation probability of the isomeric and ground states, respectively.}
\end{figure}

In order to investigate the effect of the isomeric state to the $\alpha$ decay, we compare the $\alpha$ preformation probabilities between nuclear isomers and the corresponding ground states, containing 86 favored $\alpha$ decays. For all the cases, there are two kinds of $\alpha$ transitions, i.e. isomeric states to isomeric states (i.s. to i.s.) and the corresponding ground states to ground states (g.s. to g.s.). The calculations of the logarithm of $P^*_\alpha/P_\alpha$ are plotted as a function of proton numbers of the parent nuclei in Fig.~\ref{Fig1}. $P^*_\alpha$ is the extracted $\alpha$ preformation probability for the nuclear isomer, while $P_\alpha$ is the probability for the ground state. All the experimental data of $\alpha$-decay half-lives, energies, and spin-parity, used in the calculations of this work, are taken from the latest evaluated nuclear properties table NUBASE2012~\cite{NUBASE2012}. From Fig.~\ref{Fig1}, we can see that the values of $\log_{10}P^*_\alpha/P_\alpha$ are around 0, indicating there are no obvious differences for $\alpha$ preformation probabilities between nuclear isomers and ground states to some extent, as shown in the existing researches~\cite{Qia12,*Qia11,*Qian11}. Therefore, $\alpha$ decay of both ground and isometric states can be treated in an unified form.

\begin{figure}[!htb]
\centering
\includegraphics[width=8.6cm]{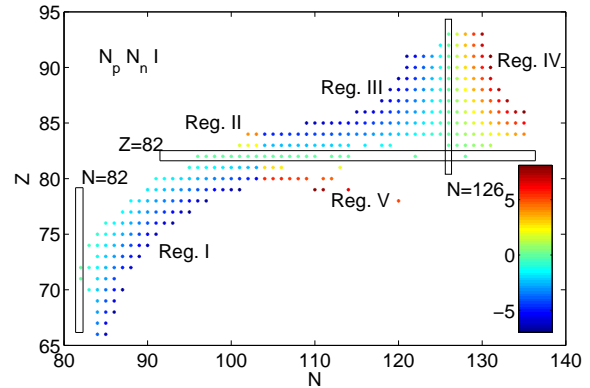}
\caption{\label{Fig2} The colormap of $N_\text{p} N_\text{n} I$ as a function of neutron numbers $N$ and proton numbers $Z$ of the parent nuclei. The rectangles denote the nuclear shell closures at $Z=82$ and $N$=82, 126, and the nuclei are divided into five regions.}
\end{figure}

Many researchers, using different models, indicated that the $\alpha$ preformation probability obviously diminishes with a small number of valence nucleons (holes)~\cite{Zha11,Qia13,Zha09}. Moreover, the valence proton-neutron interaction is an important residual interaction for mean-field approximation~\cite{Cas87,Zha00}. It has been found that $\alpha$ preformation probabilities are linearly related with the product of valence proton numbers and valence neutron numbers $N_\text{p} N_\text{n}$ for even-even nuclei around the doubly magic cores at $Z=82$, $N=82$ and at $Z=82$, $N=126$. Besides, isospin asymmetry of the $\alpha$ decay parent nucleus was also considered in the $N_\text{p} N_\text{n}$ scheme~\cite{Sei11}. $N_\text{p}$ and $N_\text{n}$ denote the valence protons (or valence proton holes) and valence neutrons (or valence neutron holes) of the parent nuclei, respectively. Then, it is interesting to see whether the $N_\text{p} N_\text{n}$ scheme works as well in odd-$A$ and doubly-odd nuclei as in even-even nuclei.

To clearly describe the $N_\text{p} N_\text{n}$ scheme, we perform candidates for $\alpha$ radioactivity on the colormap of $N_\text{p} N_\text{n} I$ as a function of neutron numbers $N$ and proton numbers $Z$ of the parent nuclei in Fig.~\ref{Fig2}. $I$ denotes the isospin asymmetry of the $\alpha$-decay parent nucleus. The quadrants in coordinate of valence nucleons are labeled from Region I to V, respectively. Because of a little number of nuclei with $\alpha$ radioactivity in Regions II and V, in this work, we focus on the $\alpha$ preformation probabilities in Regions I, III, and IV. As shown in Fig.~\ref{Fig2}, from the point of the $N_\text{p} N_\text{n}$ scheme, the $\alpha$ preformation probability in Region I decreases with increasing distance away from $\beta$ stability roughly. Contrary to the case in Region I, $\alpha$ preformation probabilities in Region III increase gradually close to the proton drip line.

The detailed numerical results of $\alpha$ transitions around the doubly magic core at $Z=82$, $N=82$ shell closures are listed in Table~\ref{Tab1}. The first four columns represent the $\alpha$ transition, decay energy, extracted $\alpha$ preformation probability, and experimental $\alpha$ decay half-life, respectively. These nuclei lied in Region I above the neutron $N=82$ shell closure and below the proton $Z=82$ shell closure. Most of them are near to the proton drip line on the chart of nuclides. The extracted $\alpha$ preformation probabilities can be evaluated due to the linear relationship between the preformation probabilities and valence nucleon product $N_\text{p} N_\text{n}$. It is believed that the slopes of $\alpha$ preformation probabilities against $N_\text{p} N_\text{n}$ are related to the average interaction of proton-neutron pairs~\cite{Zha00}. The above linear relationship implies that the influence of proton-neutron pairs on $\alpha$ clustering roughly remains constant for closed shell nuclei in the same region.

In order to gain a better insight into the agreement between experiment and theory, we study the $\alpha$ preformation probabilities from the view of the valence nucleon around the doubly magic cores at $Z=82$, $N=82$ and at $Z=82$, $N=126$ shell closures, respectively. The $\alpha$ preformation probabilities are evaluated by the linear relationship
\begin{eqnarray}\label{9}
P_\alpha=a \frac{N_\text{p} N_\text{n}}{N_0 + Z_0}+b.
\end{eqnarray}
Where $N_0$, $Z_0$ are the neighboring neutrons and protons magic numbers, respectively. The preformation probabilities can also be sized up by taking into account the isospin asymmetry~\cite{Sei11}, and it is expressed by
\begin{eqnarray}\label{10}
P_\alpha=c N_\text{p} N_\text{n} I+d.
\end{eqnarray}
The above $a$, $b$, $c$, and $d$ are free parameters. $I$ is the asymmetry between neutron and proton in parent nuclei, while the value of $I$ depends on the doubly magic core at $N_0$, $Z_0$ and the specific quadrant. Eq. (\ref{10}) suggests that $\alpha$ preformation probabilities in the fourth quadrant are larger than those in the second quadrant, while Eq. (\ref{9}) supports that the cases in each quadrant are equal. The calculated $\alpha$ decay half-lives based on Eqs.~(\ref{9}) and (\ref{10}) are listed in the last two columns of Tables~\ref{Tab1}, \ref{Tab2}, and \ref{Tab4}, labeled by $T^\text{calc1}_{1/2}$ and $T^\text{calc2}_{1/2}$, respectively.

\begingroup
\renewcommand*{\arraystretch}{1.2}
\begin{longtable*}{cccccc}
\caption{Calculations of favored $\alpha$-decay half-lives and the $\alpha$ preformation probability of odd-$A$ nuclei in Region I related to the ground and isomeric states around the doubly magic core at $Z=82$, $N=82$. The $\alpha$-decay half-lives are calculated by Eqs. (\ref{9}) and (\ref{10}), related to $\frac{N_\text{p} N_\text{n}}{N_0 + Z_0}$ and $N_\text{p} N_\text{n} I$, respectively.}
\label{Tab1} \\
\hline \hline
$\alpha$ transition & $Q_\alpha$(MeV) & $P_\alpha$ & $T^\text{expt}_{1/2}$(s) & $T^\text{calc1}_{1/2}$(s) & $T^\text{calc2}_{1/2}$(s) \\
\hline
\endfirsthead
\multicolumn{6}{c}{{\tablename\ \thetable{}. (Continued.)}} \\
\hline \hline
$\alpha$ transition & $Q_\alpha$(MeV) & $P_\alpha$ & $T^\text{expt}_{1/2}$(s) & $T^\text{calc1}_{1/2}$(s) & $T^\text{calc2}_{1/2}$(s) \\
\hline
\endhead
\hline
\endfoot
\hline \hline
\endlastfoot
$^{151}$Dy$$$\to$$^{147}$Gd$$&4.18&0.213&1.92$\times10^{4}$&1.6$\times10^{4}$&1.47$\times10^{4}$\\
$^{151}$Ho$$$\to$$^{147}$Tb$^m$&4.644&0.218&1.6$\times10^{2}$&1.65$\times10^{2}$&1.53$\times10^{2}$\\
$^{151}$Ho$^m$$\to$$^{147}$Tb$$&4.736&0.186&6.12$\times10^{1}$&5.41$\times10^{1}$&5.04$\times10^{1}$\\
$^{153}$Er$$$\to$$^{149}$Dy$$&4.802&0.245&7$\times10^{1}$&7.13$\times10^{1}$&6.82$\times10^{1}$\\
$^{153}$Tm$$$\to$$^{149}$Ho$$&5.248&0.258&1.63$\times10^{0}$&2.08$\times10^{0}$&1.99$\times10^{0}$\\
$^{153}$Tm$^m$$\to$$^{149}$Ho$^m$&5.242&0.165&2.72$\times10^{0}$&2.22$\times10^{0}$&2.12$\times10^{0}$\\
$^{155}$Tm$$$\to$$^{151}$Ho$$&4.572&0.417&2.43$\times10^{3}$&3.81$\times10^{3}$&3.73$\times10^{3}$\\
$^{155}$Yb$$$\to$$^{151}$Er$$&5.338&0.247&2.01$\times10^{0}$&2.2$\times10^{0}$&2.17$\times10^{0}$\\
$^{155}$Lu$$$\to$$^{151}$Tm$$&5.803&0.223&7.62$\times10^{-2}$&8.89$\times10^{-2}$&8.61$\times10^{-2}$\\
$^{155}$Lu$^m$$\to$$^{151}$Tm$^m$&5.73&0.185&1.82$\times10^{-1}$&1.76$\times10^{-1}$&1.71$\times10^{-1}$\\
$^{157}$Yb$$$\to$$^{153}$Er$$&4.621&0.239&7.73$\times10^{3}$&6.47$\times10^{3}$&6.44$\times10^{3}$\\
$^{157}$Lu$^m$$\to$$^{153}$Tm$$&5.128&0.186&7.98$\times10^{1}$&6.04$\times10^{1}$&6.12$\times10^{1}$\\
$^{157}$Hf$$$\to$$^{153}$Yb$$&5.885&0.168&1.34$\times10^{-1}$&1.06$\times10^{-1}$&1.07$\times10^{-1}$\\
$^{159}$Ta$$$\to$$^{155}$Lu$^m$&5.66&0.182&3.06$\times10^{0}$&2.45$\times10^{0}$&2.54$\times10^{0}$\\
$^{159}$Ta$^m$$\to$$^{155}$Lu$$&5.744&0.237&1.02$\times10^{0}$&1.07$\times10^{0}$&1.11$\times10^{0}$\\
$^{159}$W$$$\to$$^{155}$Hf$$&6.445&0.141&1$\times10^{-2}$&7.19$\times10^{-3}$&7.28$\times10^{-3}$\\
$^{159}$Re$^m$$\to$$^{155}$Ta$$&6.965&0.243&2.88$\times10^{-4}$&4.07$\times10^{-4}$&3.94$\times10^{-4}$\\
$^{161}$W$$$\to$$^{157}$Hf$$&5.915&0.238&5.6$\times10^{-1}$&5.66$\times10^{-1}$&5.96$\times10^{-1}$\\
$^{161}$Re$^m$$\to$$^{157}$Ta$^m$&6.425&0.277&1.58$\times10^{-2}$&2.12$\times10^{-2}$&2.2$\times10^{-2}$\\
$^{161}$Os$$$\to$$^{157}$W$$&7.065&0.129&6.4$\times10^{-4}$&4.55$\times10^{-4}$&4.55$\times10^{-4}$\\
$^{163}$W$$$\to$$^{159}$Hf$$&5.518&0.327&1.88$\times10^{1}$&2.23$\times10^{1}$&2.36$\times10^{1}$\\
$^{163}$Re$$$\to$$^{159}$Ta$$&6.012&0.123&1.22$\times10^{0}$&6.24$\times10^{-1}$&6.64$\times10^{-1}$\\
$^{163}$Re$^m$$\to$$^{159}$Ta$^m$&6.068&0.277&3.24$\times10^{-1}$&3.72$\times10^{-1}$&3.96$\times10^{-1}$\\
$^{163}$Os$$$\to$$^{159}$W$$&6.675&0.267&5.5$\times10^{-3}$&6.95$\times10^{-3}$&7.29$\times10^{-3}$\\
$^{165}$Re$^m$$\to$$^{161}$Ta$^m$&5.66&0.233&1.79$\times10^{1}$&1.51$\times10^{1}$&1.6$\times10^{1}$\\
$^{165}$Os$$$\to$$^{161}$W$$&6.335&0.189&1.18$\times10^{-1}$&9.27$\times10^{-2}$&9.91$\times10^{-2}$\\
$^{165}$Ir$^m$$\to$$^{161}$Re$^m$&6.885&0.307&2.31$\times10^{-3}$&3.36$\times10^{-3}$&3.53$\times10^{-3}$\\
$^{167}$Os$$$\to$$^{163}$W$$&5.985&0.293&1.64$\times10^{0}$&1.78$\times10^{0}$&1.89$\times10^{0}$\\
$^{167}$Ir$$$\to$$^{163}$Re$$&6.504&0.203&6.81$\times10^{-2}$&5.87$\times10^{-2}$&6.24$\times10^{-2}$\\
$^{167}$Ir$^m$$\to$$^{163}$Re$^m$&6.561&0.3&2.86$\times10^{-2}$&3.64$\times10^{-2}$&3.86$\times10^{-2}$\\
$^{167}$Pt$$$\to$$^{163}$Os$$&7.155&0.285&8$\times10^{-4}$&1.1$\times10^{-3}$&1.15$\times10^{-3}$\\
$^{169}$Os$$$\to$$^{165}$W$$&5.713&0.249&2.52$\times10^{1}$&2.09$\times10^{1}$&2.17$\times10^{1}$\\
$^{169}$Ir$$$\to$$^{165}$Re$$&6.141&0.393&7.85$\times10^{-1}$&1.18$\times10^{0}$&1.25$\times10^{0}$\\
$^{169}$Ir$^m$$\to$$^{165}$Re$^m$&6.266&0.252&3.9$\times10^{-1}$&3.78$\times10^{-1}$&3.98$\times10^{-1}$\\
$^{171}$Pt$$$\to$$^{167}$Os$$&6.605&0.272&5.06$\times10^{-2}$&5.6$\times10^{-2}$&5.83$\times10^{-2}$\\
$^{173}$Pt$$$\to$$^{169}$Os$$&6.358&0.237&4.45$\times10^{-1}$&3.95$\times10^{-1}$&4.04$\times10^{-1}$\\
$^{173}$Au$$$\to$$^{169}$Ir$$&6.837&0.181&2.9$\times10^{-2}$&2.33$\times10^{-2}$&2.38$\times10^{-2}$\\
$^{175}$Au$$$\to$$^{171}$Ir$$&6.575&0.192&2.16$\times10^{-1}$&1.72$\times10^{-1}$&1.73$\times10^{-1}$\\
$^{177}$Au$$$\to$$^{173}$Ir$$&6.298&0.118&3.65$\times10^{0}$&1.68$\times10^{0}$&1.65$\times10^{0}$\\
$^{177}$Tl$$$\to$$^{173}$Au$$&7.066&0.218&2.47$\times10^{-2}$&3.14$\times10^{-2}$&2.91$\times10^{-2}$\\
$^{179}$Hg$$$\to$$^{175}$Pt$$&6.351&0.376&1.91$\times10^{0}$&3.25$\times10^{0}$&3.1$\times10^{0}$\\
$^{183}$Hg$$$\to$$^{179}$Pt$$&6.038&0.141&8.04$\times10^{1}$&4.71$\times10^{1}$&4.32$\times10^{1}$\\
$^{183}$Tl$$$\to$$^{179}$Au$$&5.977&0.192&3.45$\times10^{2}$&3.56$\times10^{2}$&3.21$\times10^{2}$\\
$^{185}$Pb$^m$$\to$$^{181}$Hg$^m$&6.555&0.089&8.15$\times10^{0}$&5.28$\times10^{0}$&4.46$\times10^{0}$\\
\end{longtable*}
\endgroup

In Table~\ref{Tab2}, we present the theoretical results of $\alpha$ transitions for nuclei around the doubly magic core at $Z=82$, $N=126$ shell closures. These nuclei are divided into two groups, lied in the first quadrant (Region IV) and the second quadrant (Region III) of the coordinate of valence nucleons, respectively. The nuclei in Region III, whose $N_\text{p} N_\text{n}$ are negative, involve valence protons and valence neutron holes. While $N_\text{p} N_\text{n}$ of nuclei in Region IV are positive.

\begin{figure}[!htb]
\centering
\includegraphics[width=8.6cm]{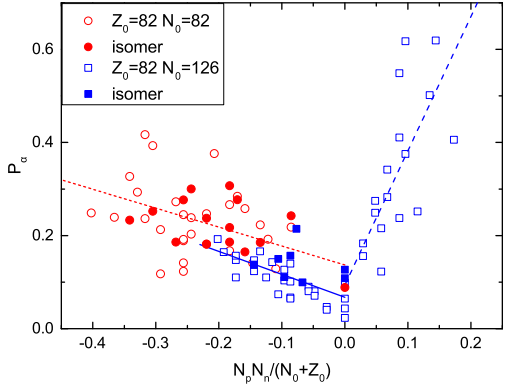}
\caption{\label{Fig3} The $\alpha$ preformation probabilities as a function of $\frac{N_\text{p} N_\text{n}}{N_0 + Z_0}$, where $N_\text{p}$, $N_\text{n}$ represent valence proton numbers and valence neutron numbers of the parent nucleus, respectively, and $N_0 (Z_0)$ express neutron (proton) magic numbers. The blue dashed and solid lines denote the fit of nuclei in Regions IV and III, respectively. The red short dashed line denotes the fit of nuclei in Region I.}
\end{figure}

\begin{figure}[!htb]
\centering
\includegraphics[width=8.6cm]{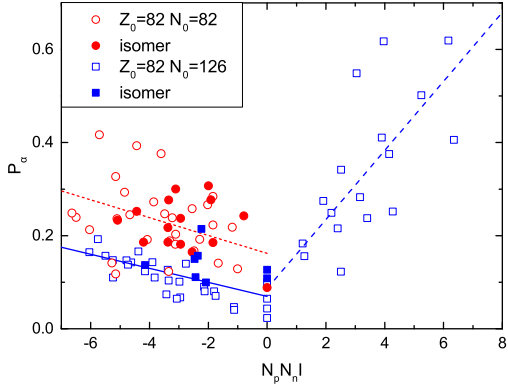}
\caption{\label{Fig4} Same as Figure~\ref{Fig3}, but as a function of the product of valence proton numbers and valence neutron numbers and isospin asymmetry $N_\text{p} N_\text{n} I$.}
\end{figure}

The $\alpha$ preformation probabilities $P_\alpha$ are plotted as the function of $\frac{N_\text{p} N_\text{n}}{N_0 + Z_0}$ in Figure~\ref{Fig3}, and of $N_\text{p} N_\text{n} I$ in Figure~\ref{Fig4}, respectively. The open red circles and blue squares denote nuclei in ground states around the doubly magic cores at $Z=82$, $N=82$ and at $Z=82$, $N=126$ shell closures, respectively. The filled ones correspond to nuclear isomers. As is shown, there is very much similarity in ground and isomeric states for $\alpha$ decay. The lines denote the predictions of Eqs. (\ref{9}) and (\ref{10}). The results indicate the average residual proton-neutron interaction is different for closed shell nuclei around distinct shell closures. Clearly, the linear relationship as a function of $\frac{N_\text{p} N_\text{n}}{N_0 + Z_0}$ in Figure~\ref{Fig3} is slightly better than that of $N_\text{p} N_\text{n} I$ in Figure~\ref{Fig4}. Besides the predictions in Regions III and IV are better than those in Region I. Maybe it is because the doubly magic core at $Z=82$, $N=82$ is unbound, and the nucleons in the core play an essential role on the $\alpha$ preformation probability.

\begingroup
\renewcommand*{\arraystretch}{1.2}
\begin{longtable*}{cccccc}
\caption{Same as Table I, but for favored $\alpha$ decay of odd-$A$ nuclei around the doubly magic core at $Z=82$, $N=126$.}
\label{Tab2} \\
\hline \hline
$\alpha$ transition & $Q_\alpha$(MeV) & $P_\alpha$ & $T^\text{expt}_{1/2}$(s) & $T^\text{calc1}_{1/2}$(s) & $T^\text{calc2}_{1/2}$(s) \\
\hline
\endfirsthead
\multicolumn{6}{c}{{\tablename\ \thetable{}. (Continued.)}} \\
\hline \hline
$\alpha$ transition & $Q_\alpha$(MeV) & $P_\alpha$ & $T^\text{expt}_{1/2}$(s) & $T^\text{calc1}_{1/2}$(s) & $T^\text{calc2}_{1/2}$(s) \\
\hline
\endhead
\hline
\endfoot
\hline \hline
\endlastfoot
\multicolumn{6}{c}{nuclei in Region III}\\
$^{187}$Pb$^m$$\to$$^{183}$Hg$^m$&6.208&0.109&1.52$\times10^{2}$&1.76$\times10^{2}$&1.92$\times10^{2}$\\
$^{187}$Bi$^m$$\to$$^{183}$Tl$$&7.887&0.15&3.7$\times10^{-4}$&4.64$\times10^{-4}$&5.21$\times10^{-4}$\\
$^{189}$Bi$^m$$\to$$^{185}$Tl$$&7.452&0.111&9.79$\times10^{-3}$&9.48$\times10^{-3}$&1.03$\times10^{-2}$\\
$^{191}$Pb$^m$$\to$$^{187}$Hg$^m$&5.404&0.127&6.55$\times10^{5}$&8.82$\times10^{5}$&9.62$\times10^{5}$\\
$^{191}$Bi$^m$$\to$$^{187}$Tl$$&7.018&0.157&1.82$\times10^{-1}$&2.6$\times10^{-1}$&2.73$\times10^{-1}$\\
$^{193}$Bi$^m$$\to$$^{189}$Tl$$&6.613&0.215&3.81$\times10^{0}$&7.75$\times10^{0}$&7.93$\times10^{0}$\\
$^{195}$Bi$^m$$\to$$^{191}$Tl$$&6.232&0.099&2.64$\times10^{2}$&2.6$\times10^{2}$&2.6$\times10^{2}$\\
$^{195}$Po$$$\to$$^{191}$Pb$$&6.755&0.124&4.93$\times10^{0}$&4.4$\times10^{0}$&4.62$\times10^{0}$\\
$^{195}$Po$^m$$\to$$^{191}$Pb$^m$&6.84&0.137&2.13$\times10^{0}$&2.1$\times10^{0}$&2.21$\times10^{0}$\\
$^{197}$Po$$$\to$$^{193}$Pb$$&6.405&0.11&1.22$\times10^{2}$&1.04$\times10^{2}$&1.06$\times10^{2}$\\
$^{197}$At$$$\to$$^{193}$Bi$$&7.108&0.193&4.04$\times10^{-1}$&4.66$\times10^{-1}$&4.98$\times10^{-1}$\\
$^{199}$Po$$$\to$$^{195}$Pb$$&6.074&0.074&4.38$\times10^{3}$&2.71$\times10^{3}$&2.67$\times10^{3}$\\
$^{199}$At$$$\to$$^{195}$Bi$$&6.778&0.148&7.89$\times10^{0}$&7.62$\times10^{0}$&7.84$\times10^{0}$\\
$^{201}$Po$$$\to$$^{197}$Pb$$&5.799&0.067&8.28$\times10^{4}$&5.06$\times10^{4}$&4.89$\times10^{4}$\\
$^{201}$At$$$\to$$^{197}$Bi$$&6.473&0.143&1.2$\times10^{2}$&1.24$\times10^{2}$&1.23$\times10^{2}$\\
$^{203}$At$$$\to$$^{199}$Bi$$&6.209&0.143&1.43$\times10^{3}$&1.65$\times10^{3}$&1.6$\times10^{3}$\\
$^{203}$Rn$$$\to$$^{199}$Po$$&6.63&0.157&6.67$\times10^{1}$&6.85$\times10^{1}$&6.88$\times10^{1}$\\
$^{205}$Po$$$\to$$^{201}$Pb$$&5.325&0.081&1.57$\times10^{7}$&1.38$\times10^{7}$&1.31$\times10^{7}$\\
$^{205}$At$$$\to$$^{201}$Bi$$&6.02&0.065&2.03$\times10^{4}$&1.19$\times10^{4}$&1.13$\times10^{4}$\\
$^{205}$Fr$$$\to$$^{201}$At$$&7.054&0.165&3.82$\times10^{0}$&3.88$\times10^{0}$&3.92$\times10^{0}$\\
$^{207}$Po$$$\to$$^{203}$Pb$$&5.216&0.047&9.95$\times10^{7}$&5.69$\times10^{7}$&5.38$\times10^{7}$\\
$^{207}$At$$$\to$$^{203}$Bi$$&5.872&0.09&6.52$\times10^{4}$&6.14$\times10^{4}$&5.79$\times10^{4}$\\
$^{207}$Rn$$$\to$$^{203}$Po$$&6.251&0.127&2.64$\times10^{3}$&2.91$\times10^{3}$&2.78$\times10^{3}$\\
$^{207}$Fr$$$\to$$^{203}$At$$&6.894&0.147&1.56$\times10^{1}$&1.65$\times10^{1}$&1.62$\times10^{1}$\\
$^{209}$At$$$\to$$^{205}$Bi$$&5.757&0.041&4.75$\times10^{5}$&2.36$\times10^{5}$&2.24$\times10^{5}$\\
$^{209}$Rn$$$\to$$^{205}$Po$$&6.155&0.08&1.01$\times10^{4}$&8.44$\times10^{3}$&7.99$\times10^{3}$\\
$^{209}$Fr$$$\to$$^{205}$At$$&6.777&0.104&5.62$\times10^{1}$&5.06$\times10^{1}$&4.85$\times10^{1}$\\
$^{209}$Ra$$$\to$$^{205}$Rn$$&7.135&0.138&5.24$\times10^{0}$&5.21$\times10^{0}$&5.12$\times10^{0}$\\
$^{211}$At$$$\to$$^{207}$Bi$$&5.982&0.023&6.21$\times10^{4}$&1.54$\times10^{4}$&1.68$\times10^{4}$\\
$^{211}$Fr$$$\to$$^{207}$At$$&6.662&0.071&2.14$\times10^{2}$&1.66$\times10^{2}$&1.58$\times10^{2}$\\
$^{211}$Ra$$$\to$$^{207}$Rn$$&7.042&0.101&1.42$\times10^{1}$&1.31$\times10^{1}$&1.26$\times10^{1}$\\
$^{211}$Ac$$$\to$$^{207}$Fr$$&7.619&0.166&2.13$\times10^{-1}$&2.64$\times10^{-1}$&2.61$\times10^{-1}$\\
$^{213}$Fr$$$\to$$^{209}$At$$&6.904&0.044&3.5$\times10^{1}$&1.62$\times10^{1}$&1.77$\times10^{1}$\\
$^{213}$Pa$$$\to$$^{209}$Ac$$&8.395&0.11&7$\times10^{-3}$&5.05$\times10^{-3}$&5.2$\times10^{-3}$\\
$^{215}$Ac$$$\to$$^{211}$Fr$$&7.746&0.064&1.7$\times10^{-1}$&1.16$\times10^{-1}$&1.27$\times10^{-1}$\\
$^{215}$Pa$$$\to$$^{211}$Ac$$&8.245&0.14&1.4$\times10^{-2}$&1.78$\times10^{-2}$&1.76$\times10^{-2}$\\
$^{217}$Pa$$$\to$$^{213}$Ac$$&8.485&0.099&3.48$\times10^{-3}$&3.66$\times10^{-3}$&3.99$\times10^{-3}$\\
\multicolumn{6}{c}{nuclei in Region IV}\\
$^{213}$Po$$$\to$$^{209}$Pb$$&8.536&0.156&3.72$\times10^{-6}$&3.27$\times10^{-6}$&3.22$\times10^{-6}$\\
$^{213}$At$$$\to$$^{209}$Bi$$&9.254&0.183&1.25$\times10^{-7}$&1.29$\times10^{-7}$&1.3$\times10^{-7}$\\
$^{215}$Po$$$\to$$^{211}$Pb$$&7.526&0.249&1.78$\times10^{-3}$&1.91$\times10^{-3}$&1.79$\times10^{-3}$\\
$^{215}$At$$$\to$$^{211}$Bi$$&8.178&0.123&1$\times10^{-4}$&4.72$\times10^{-5}$&4.5$\times10^{-5}$\\
$^{215}$Rn$$$\to$$^{211}$Po$$&8.839&0.216&2.3$\times10^{-6}$&1.91$\times10^{-6}$&1.88$\times10^{-6}$\\
$^{215}$Fr$$$\to$$^{211}$At$$&9.54&0.275&8.59$\times10^{-8}$&1.02$\times10^{-7}$&1.04$\times10^{-7}$\\
$^{217}$Po$$$\to$$^{213}$Pb$$&6.662&0.283&1.59$\times10^{0}$&1.57$\times10^{0}$&1.4$\times10^{0}$\\
$^{217}$At$$$\to$$^{213}$Bi$$&7.201&0.411&3.23$\times10^{-2}$&3.87$\times10^{-2}$&3.53$\times10^{-2}$\\
$^{217}$Rn$$$\to$$^{213}$Po$$&7.887&0.375&5.4$\times10^{-4}$&5.47$\times10^{-4}$&5.15$\times10^{-4}$\\
$^{217}$Fr$$$\to$$^{213}$At$$&8.469&0.618&1.68$\times10^{-5}$&2.8$\times10^{-5}$&2.73$\times10^{-5}$\\
$^{217}$Ra$$$\to$$^{213}$Rn$$&9.161&0.238&1.63$\times10^{-6}$&1.13$\times10^{-6}$&1.14$\times10^{-6}$\\
$^{217}$Ac$$$\to$$^{213}$Fr$$&9.832&0.342&6.9$\times10^{-8}$&8.19$\times10^{-8}$&8.64$\times10^{-8}$\\
$^{219}$Fr$$$\to$$^{215}$At$$&7.449&0.619&2$\times10^{-2}$&2.43$\times10^{-2}$&2.28$\times10^{-2}$\\
$^{219}$Ac$$$\to$$^{215}$Fr$$&8.827&0.502&1.18$\times10^{-5}$&1.23$\times10^{-5}$&1.25$\times10^{-5}$\\
$^{219}$Th$$$\to$$^{215}$Ra$$&9.511&0.252&1.05$\times10^{-6}$&6.2$\times10^{-7}$&6.55$\times10^{-7}$\\
$^{219}$Pa$$$\to$$^{215}$Ac$$&10.084&0.549&5.3$\times10^{-8}$&8.48$\times10^{-8}$&9.33$\times10^{-8}$\\
$^{221}$Pa$$$\to$$^{217}$Ac$$&9.251&0.406&5.9$\times10^{-6}$&4.04$\times10^{-6}$&4.29$\times10^{-6}$\\
\end{longtable*}
\endgroup

The parameters $a$, $b$, $c$, and $d$ of Eqs. (\ref{9}) and (\ref{10}) are presented in Table~\ref{Tab3}. The $\alpha$ transitions of ground states and nuclear isomers are treated in a unified way. Because the value of isospin $I$ changes little, the values of both $b$ and $d$ for nuclei in distinct regions given by Eq. (\ref{9}) are nearly equal to those given by Eq. (\ref{10}). And the calculations involve favored $\alpha$ decay of both odd-$A$ and doubly-odd nuclei.

\begin{table}
\caption{\label{Tab3}The parameters of Eqs. (\ref{9}) and (\ref{10}) that show $\alpha$ preformation probabilities are linearly related to the valence proton-neutron interaction.}
\begin{ruledtabular}
\begin{tabular}{ccccc}
Region & \multicolumn{2}{c}{Eq. (\ref{9})} & \multicolumn{2}{c}{Eq. (\ref{10})} \\ \cline{2-3} \cline{4-5}
 & $a$ & $b$ & $c$ & $d$ \\ \hline
 & \multicolumn{4}{c}{odd-$A$ nuclei} \\
I & -1.162 & 0.137 & -0.055 & 0.162 \\
III & -1.409 & 0.068 & -0.043 & 0.069 \\
IV & 8.230 & 0.094 & 0.212 & 0.086 \\
 & \multicolumn{4}{c}{doubly-odd nuclei} \\
I & -1.831 & 0.093 & -0.011 & 0.154 \\
III & -3.477 & -0.006 & -0.128 & -0.021 \\
IV & 6.676 & 0.105 & 0.112 & 0.137 \\
\end{tabular}
\end{ruledtabular}
\end{table}

\begin{figure}[!htb]
\centering
\includegraphics[width=8.6cm]{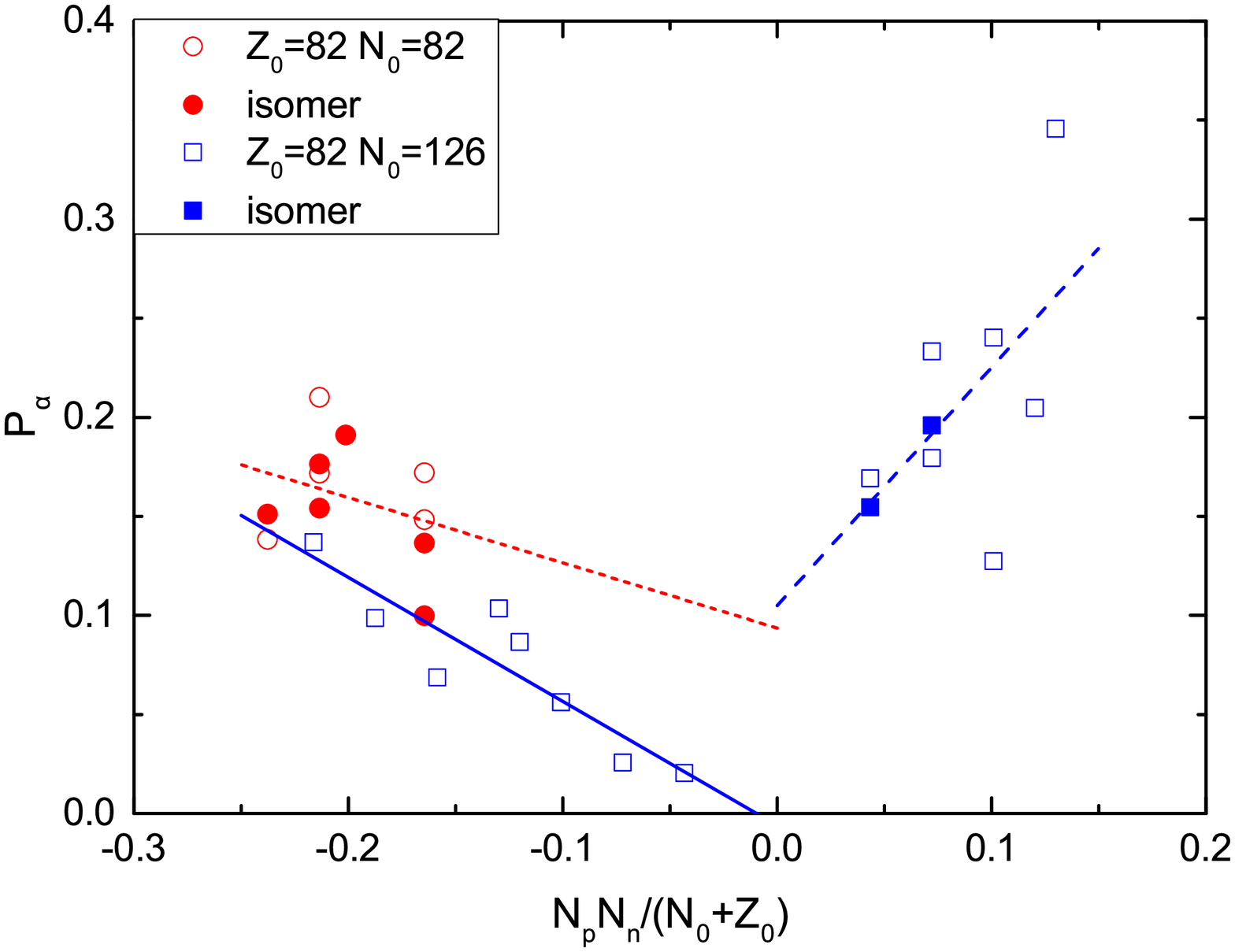}
\caption{\label{Fig5} Same as Figue 3, but for doubly-odd nuclei as a function of $\frac{N_\text{p} N_\text{n}}{N_0 + Z_0}$.}
\end{figure}

\begin{figure}[!htb]
\centering
\includegraphics[width=9cm]{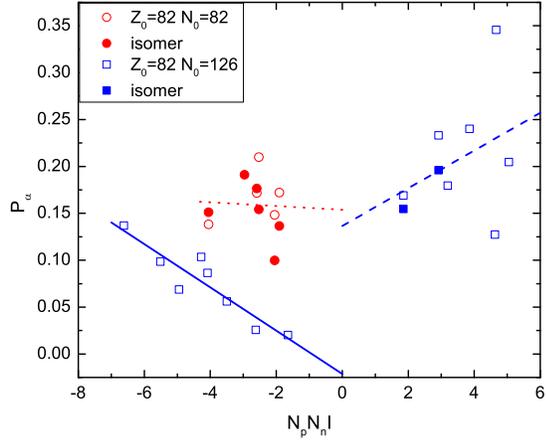}
\caption{\label{Fig6} Same as Figue 3, but for doubly-odd nuclei as a function of $N_\text{p} N_\text{n}I$.}
\end{figure}

For the favored $\alpha$ decay of doubly-odd nuclei, the linear relationship still exists as shown in Figs.~\ref{Fig5} and \ref{Fig6}. The calculated $\alpha$ decay half-lives are obtained by Eqs. (\ref{9}) and (\ref{10}) with parameters presented in Table~\ref{Tab3}, and the detailed results of doubly-odd nuclei are listed in Table~\ref{Tab4}. Although there are much less $\alpha$-decay cases for closed shell doubly-odd nuclei, the listed data still imply the varying trend of $\alpha$ preformation probability due to the valence proton-neutron interaction. The $\alpha$ preformation probability almost increase linearly with the increases of the valence proton-neutron interaction.

\begingroup
\renewcommand*{\arraystretch}{1.2}
\begin{longtable*}{cccccc}
\caption{Same as Table I, but for favored $\alpha$ decay of doubly-odd nuclei.}
\label{Tab4} \\
\hline \hline
$\alpha$ transition & $Q_\alpha$(MeV) & $P_\alpha$ & $T^\text{expt}_{1/2}$(s) & $T^\text{calc1}_{1/2}$(s) & $T^\text{calc2}_{1/2}$(s) \\
\hline
\endfirsthead
\multicolumn{6}{c}{{\tablename\ \thetable{}. (Continued.)}} \\
\hline \hline
$\alpha$ transition & $Q_\alpha$(MeV) & $P_\alpha$ & $T^\text{expt}_{1/2}$(s) & $T^\text{calc1}_{1/2}$(s) & $T^\text{calc2}_{1/2}$(s) \\
\hline
\endhead
\hline
\endfoot
\hline \hline
\endlastfoot
\multicolumn{6}{c}{doubly-odd nuclei in Region I around the doubly magic core at Z=82, N=82}\\
$^{154}$Tm$\to$$^{150}$Ho&5.094&0.138&1.5$\times10^{1}$&1.21$\times10^{1}$&1.28$\times10^{1}$\\
$^{154}$Tm$^m$$\to$$^{150}$Ho$^m$&5.175&0.151&5.69$\times10^{0}$&5.01$\times10^{0}$&5.32$\times10^{0}$\\
$^{156}$Lu$^m$$\to$$^{152}$Tm$^m$&5.705&0.191&2.11$\times10^{-1}$&2.52$\times10^{-1}$&2.52$\times10^{-1}$\\
$^{158}$Ta$\to$$^{154}$Lu&6.125&0.148&5.1$\times10^{-2}$&5.13$\times10^{-2}$&4.8$\times10^{-2}$\\
$^{158}$Ta$^m$$\to$$^{154}$Lu$^m$&6.205&0.1&3.79$\times10^{-2}$&2.56$\times10^{-2}$&2.39$\times10^{-2}$\\
$^{162}$Re$\to$$^{158}$Ta&6.245&0.172&1.14$\times10^{-1}$&1.19$\times10^{-1}$&1.23$\times10^{-1}$\\
$^{162}$Re$^m$$\to$$^{158}$Ta$^m$&6.275&0.177&8.46$\times10^{-2}$&9.12$\times10^{-2}$&9.4$\times10^{-2}$\\
$^{166}$Ir$\to$$^{162}$Re&6.725&0.21&1.13$\times10^{-2}$&1.45$\times10^{-2}$&1.49$\times10^{-2}$\\
$^{166}$Ir$^m$$\to$$^{162}$Re$^m$&6.725&0.154&1.54$\times10^{-2}$&1.45$\times10^{-2}$&1.49$\times10^{-2}$\\
$^{170}$Au$\to$$^{166}$Ir&7.175&0.172&2.64$\times10^{-3}$&3.07$\times10^{-3}$&2.88$\times10^{-3}$\\
$^{170}$Au$^m$$\to$$^{166}$Ir$^m$&7.285&0.137&1.48$\times10^{-3}$&1.36$\times10^{-3}$&1.28$\times10^{-3}$\\
\multicolumn{6}{c}{doubly-odd nuclei in Region III around the doubly magic core at Z=82, N=126}\\
$^{198}$At$\to$$^{194}$Bi&6.895&0.099&4.48$\times10^{0}$&3.97$\times10^{0}$&4.16$\times10^{0}$\\
$^{200}$At$\to$$^{196}$Bi&6.596&0.069&8.31$\times10^{1}$&6.12$\times10^{1}$&6.14$\times10^{1}$\\
$^{202}$At$\to$$^{198}$Bi&6.353&0.104&4.97$\times10^{2}$&6.84$\times10^{2}$&6.65$\times10^{2}$\\
$^{204}$At$\to$$^{200}$Bi&6.071&0.056&1.44$\times10^{4}$&1.41$\times10^{4}$&1.36$\times10^{4}$\\
$^{204}$Fr$\to$$^{200}$At&7.17&0.137&1.82$\times10^{0}$&1.93$\times10^{0}$&1.9$\times10^{0}$\\
$^{206}$At$\to$$^{202}$Bi&5.887&0.026&2.04$\times10^{5}$&1.34$\times10^{5}$&1.34$\times10^{5}$\\
$^{208}$At$\to$$^{204}$Bi&5.751&0.02&1.07$\times10^{6}$&1.03$\times10^{6}$&1.3$\times10^{6}$\\
$^{208}$Fr$\to$$^{204}$At&6.784&0.086&6.64$\times10^{1}$&8.3$\times10^{1}$&7.86$\times10^{1}$\\
\multicolumn{6}{c}{doubly-odd nuclei in Region IV around the doubly magic core at Z=82, N=126}\\
$^{214}$At$\to$$^{210}$Bi&8.988&0.169&5.58$\times10^{-7}$&6.01$\times10^{-7}$&5.43$\times10^{-7}$\\
$^{214}$At$^n$$\to$$^{210}$Bi$^m$&8.949&0.155&7.6$\times10^{-7}$&7.48$\times10^{-7}$&6.76$\times10^{-7}$\\
$^{216}$At$\to$$^{212}$Bi&7.95&0.18&3$\times10^{-4}$&2.81$\times10^{-4}$&2.68$\times10^{-4}$\\
$^{216}$Fr$\to$$^{212}$At&9.174&0.233&7$\times10^{-7}$&8.52$\times10^{-7}$&8.36$\times10^{-7}$\\
$^{216}$Fr$^m$$\to$$^{212}$At$^m$&9.17&0.196&8.5$\times10^{-7}$&8.69$\times10^{-7}$&8.54$\times10^{-7}$\\
$^{218}$At$\to$$^{214}$Bi&6.874&0.127&1.5$\times10^{0}$&8.44$\times10^{-1}$&8.32$\times10^{-1}$\\
$^{218}$Fr$\to$$^{214}$At&8.013&0.205&1$\times10^{-3}$&8.21$\times10^{-4}$&8.6$\times10^{-4}$\\
$^{218}$Ac$\to$$^{214}$Fr&9.373&0.24&1.08$\times10^{-6}$&1.15$\times10^{-6}$&1.21$\times10^{-6}$\\
$^{220}$Pa$\to$$^{216}$Ac&9.65&0.346&7.8$\times10^{-7}$&1.03$\times10^{-6}$&1.17$\times10^{-6}$\\
\end{longtable*}
\endgroup

For the case of unfavored $\alpha$ decay, the spin-parity state of the parent nucleus is different from that of the daughter nucleus. It is well known that the centrifugal potential reduces the $\alpha$-decay width. In addition, the $\alpha$ preformation probability may also be affected by the changes of nuclear structure configurations. We will make more detailed theoretical investigations in the future.

\section{Summary}
In summary, we have performed a systematic study of $\alpha$ decay for closed shell odd-$A$ and doubly-odd nuclei related to ground and isomeric states around the doubly magic cores at $Z=82$, $N=82$ and at $Z=82$, $N=126$ shell closures, respectively, within a two-potential approach. The $\alpha$-decay widths are calculated by using the semiclassical WKB method along with the isospin dependent nuclear potential. The $\alpha$ preformation probabilities are evaluated by the linear relationships of $N_\text{p} N_\text{n}$ and $N_\text{p} N_\text{n}I$, considering the shell effect and residual valence proton-neutron interaction. In order to clearly show the different shell closures, the $\alpha$ decay candidates are divided into five regions in the $N_\text{p} N_\text{n}$ or $N_\text{p} N_\text{n}I$ scheme. It is found that closed shell odd-$A$ and doubly-odd nuclei also satisfy the linear relationships, and the calculated half-lives agree well with the experimental data within a factor of 2 or better. The results show that the residual proton-neutron interaction is essential to evaluate the $\alpha$ preformation probabilities of even-even nuclei as well as odd-$A$ and doubly odd nuclei. This work will provide theoretical supports for future experiments.

\begin{acknowledgments}

This work is supported in part by the National Natural Science Foundation of China (Grant No.11205083), the construct program of the key discipline in hunan province, the Research Foundation of Education Bureau of Hunan Province,China (Grant No.15A159),the Natural Science Foundation of Hunan Province,China (Grant No.2015JJ3103),the Innovation Group of Nuclear and Particle Physics in USC, Hunan Provincial Innovation Foundation For Postgraduate (Grant No.CX2015B398).

\end{acknowledgments}

\end{document}